\documentstyle[12pt]{article}
\begin{document}
\title{Chaos in Anisotropic Pre-Inflationary Universes}
\author{ H. P. de Oliveira, I. Dami\~{a}o Soares and T. J. Stuchi}
\maketitle
\label{address}
\begin{abstract}

We study the dynamics of anisotropic Bianchi type-IX models with matter
and cosmological constant. The models can be thought as describing the
role of anisotropy in the early stages of inflation, where the
cosmological constant $\Lambda$ plays the role of the vacuum energy of
the inflaton field. The concurrence of the cosmological constant and
anisotropy are sufficient to produce a chaotic dynamics in the
gravitational degrees of freedom, connected to the presence of a
critical point of saddle-center type in the phase space of the system.
In the neighborhood of the saddle-center, the phase space presents the
structure of cylinders emanating from unstable periodic orbits. The
non-integrability of the system implies that the extension of the
cylinders away from this neighborhood has a complicated structure
arising from their transversal crossings, resulting in a chaotic
dynamics. The invariant character of chaos is guaranteed by the topology
of cylinders. The model also presents a strong asymptotic de Sitter
attractor but the way out from the initial singularity to the
inflationary phase is completely chaotic. For a large set of initial
conditions, even with very small anisotropy, the gravitational degrees
of freedom oscillate a long time in the neighborhood of the
saddle-center before recollapsing or escaping to the de Sitter phase.
These oscillations may provide a resonance mechanism for amplification
of specific wavelengths of inhomogeneous fluctuations in the models. A
geometrical interpretation is given for Wald's inequality in terms of
invariant tori and their destruction by increasing values of the
cosmological constant.

\end{abstract}

\label{PACS number:47.75+f}

\vfill\eject

\section{Introduction}

One of the cornerstones in the paradigm of inflation\cite{inflation} is
the presence of the cosmological constant, arising as the vacuum energy
of the inflaton field. The cosmological constant plays a fundamental
role in the gravitational dynamics of the inflationary model, by
inducing an exponential expansion of the scales of the model toward the
de Sitter configuration.  This asymptotic approach to the de Sitter
solution constitutes the basis of the so-called cosmic no-hair
conjecture. In the realm of homogeneous cosmologies, Wald\cite{wald}
showed that all initially expanding Bianchi cosmological models with
positive cosmological constant, except Bianchi IX, evolve towards the de
Sitter configuration. Bianchi IX models demand further that the absolute
value of the cosmological constant be sufficiently large compared with
spatial curvature terms. For the more general case of inhomogeneous
models, Starobinskii\cite{star} showed that they do inflate if a
positive cosmological constant is present. This crucial aspect of the
cosmological constant has been sufficiently emphasized in the
literature, and many authors have examined its role in producing
non-trivial dynamics in the early stages of inflation. In particular,
Calzetta and El Hasi\cite{calza}, and Cornish and Levin\cite{cornish}
exhibited chaotic behavior in the dynamics of Friedmann-Robertson-Walker
models with a cosmological constant term and scalar fields conformally
and/or minimally coupled to the curvature. Due to this feature of the
dynamics, small fluctuations in initial conditions of the model preclude
or induce the universe to inflate.  The Hamiltonian dynamical system
originating from the field equations is complex. In Ref. [4], the
effective degrees of freedom of the model are the scale factor and a
conformally coupled scalar field. Such a scalar field is interpreted as
a radiation field, which is later assumed to gain mass by the presence
of the inflaton field. In Cornish's work, the degrees of freedom are the
scale factor and two minimally and/or conformally coupled scalar fields.
In both cases, the cosmological constant is responsible for the
existence of a critical point $S$ in the finite region of the phase
space of the model. $S$ is a pure saddle point\cite{wiggins} and the
separatrices emanating from $S$ are wholly contained in an invariant
plane of the dynamical system, corresponding to the gravitational degree
of freedom only. The separatrices connect $S$ to other critical points,
and are denoted heteroclinic\cite{berry}. These connections are known to
be highly unstable, and their breaking, due to the perturbations come
from the coupling of the gravitational variable with the scalar fields,
are basically responsible for the chaotic dynamics referred to above.
This chaotic behavior is consequence of the so-called Poincar\'{e} 
homoclinic phenomena in dynamical systems\cite{ivano}.

There is however another important feature in the pre-inflationary
dynamics, arising from the presence of a positive cosmological constant,
whenever anisotropy is also present even in the form of small
perturbations. This point has not been emphasized yet in the literature
of inflation and will be object of our paper. The degrees of freedom of
the system are taken basically in the gravitational sector and, for
simplicity we restrict ourselves to two distinct scale factors. The
conjunction of the cosmological constant and anisotropy implies the
existence of a critical point $E$ in phase space, identified as a
saddle-center\cite{wiggins}. As a consequence, we have a wealthy
dynamics based on structures as homoclinic cylinders which emanate from
unstable periodic orbits that exist in a neighborhood of $E$. Analogous
to the breaking and crossing of heteroclinic curves, for instance, these
cylinders will cross each other in non-integrable cases producing a
chaotic dynamics. As we shall discuss, these structures will constitute
an invariant characterization of chaos\cite{matsas} in the models. We
will also discuss their implications for the occurrence or not of
inflation, as well as for the physics in the early stages of inflation.
The paper is organized as follows. In Sec. II we establish the
Hamiltonian and the basic characteristics of the model. Sec. III is
devoted to present the cylindrical topology near the saddle-center which
will be very useful to understand the dynamics of orbits on the phase
space. The numerical evidence for a physically relevant chaotic behavior
is showed in Sec. IV, whereas in Sec. V an interesting connection
between the break up of torus and the Wald's analysis is discussed.
Finally, in Sec. VI, we conclude and trace some perspectives of the
present work.

\section{The Dynamics of the Model}

We consider anisotropic Bianchi IX cosmological models characterized by
two scale functions $A(t)$ and $B(t)$ with the line element

\begin{equation}
d\,s^2=d\,t^2-A^2(t)\,(w^1)^2-B^2(t)\,[(w^2)^2+(w^3)^2].
\end{equation}

\noindent Here $t$ is the cosmological time and $(w^1, w^2, w^3)$ are
invariant 1-forms for Bianchi IX models\cite{bianchi}. The matter
content is a assumed to be a perfect fluid with velocity field
$\delta_0^{\mu}$ in the comoving coordinate system used plus a
cosmological constant $\Lambda$.  The cosmological constant term is
interpreted as arising from the vacuum energy of the inflaton field such
that our models may provide a description for a pre-inflationary
anisotropic stage of the universe. The energy-momentum tensor of the
fluid is described by

\begin{equation}
T^{\mu\nu} = (\rho + p)\,\delta^\mu_0\,\delta^\nu_0  - p\,g^{\mu\nu}
\end{equation}

\noindent where $\rho$ and $p$ are the energy density and pressure,
respectively. We assume the equation of state $p=\gamma\,\rho$, $0 \leq
\gamma \leq 1$. For sake of simplicity, we restrict ourselves to the
case of dust ($\gamma=0$). Distinct features arising in the cases of
$\gamma \neq 0$ will be discussed in the conclusions.

Einstein's field equations\cite{Landau}\footnote{We assume here
$8\,\pi\,G=c=1$}

\begin{equation}
G^{\mu\nu}-\Lambda\,g^{\mu\nu}=T^{\mu\nu}
\end{equation}

\noindent for Eqs. (1) and (2) can be obtained from the Hamiltonian
constraint

\begin{equation}
H(A, B, P_A, P_B) = \frac{P_A\,P_B}{4\,B} - \frac{A\,P_A^2}{8\,B^2} + 2\,A
- \frac{A^3}{2\,B^2} - 2\,\Lambda\,A\,B^2 - E_0 = 0.
\end{equation}

\noindent where $P_A$ and $P_B$ are the momenta canonically conjugated
to $A$ and $B$, respectively, and $E_0$ is a constant proportional to
the total energy of the models. It also occurs as the first integral of
the Bianchi identities, $2\,\rho\,A\,B^2=E_0$. The full dynamics is
governed by the Hamilton's equations

\begin{equation}
\left\{
\begin{array}{ll}
\dot{A} = \frac{\partial H}{\partial P_A} = \frac{P_B}{4\,B} -
\frac{A\,P_A}{4\,B^2} \\
\dot{B} = \frac{\partial H}{\partial P_B} = \frac{P_A}{4\,B} \\
\dot{P_A} = -\frac{\partial H}{\partial A} = \frac{P_A^2}{8\,B^2} - 2 +
\frac{3\,A^2}{2\,B^2} + 2\,\Lambda\,B^2 \\
\dot{P_B} = -\frac{\partial H}{\partial B} = \frac{P_A\,P_B}{4\,B^2} -
\frac{A\,P_A^2}{4\,B^3} - \frac{A^3}{B^3} + 4\,\Lambda\,A\,B\
\end{array}
\right. \
\end{equation}

\noindent The dynamical system (5) has one critical point $E$ in the
finite region of the phase space whose coordinates are

\begin{equation}
E: \, \, A_0 = B_0 = \frac{1}{\sqrt{4\,\Lambda}}; P_A = P_B = 0,
\end{equation}

\noindent with associated energy $E_0 = E_{cr}= \frac{1}{\sqrt{4\,\Lambda}}$.
This critical point represents the static Einstein universe. Linearizing (5)
about the critical point $E$, we can show that the constant matrix
determining the linear system about $E$ has the four eigenvalues

\begin{equation}
\lambda_{1,2} = \pm\,\frac{1}{2\,E_{cr}}; \lambda_{3,4} =
\pm\,\frac{i\,\sqrt{2}}{E_{cr}},
\end{equation}

\noindent which characterizes the $E$ as a saddle-center. The system (5)
has also a degenerate critical point\cite{bogoy} at $A=B=0$,
$P_A=P_B=0$. A straightforward analysis at the infinity of the phase
space under consideration shows that it has two critical points at this
region, corresponding to the de Sitter solution, one acting as an
attractor (stable de Sitter configuration) and the other as a repeller
(unstable de Sitter configuration). The scale factors $A$ and $B$
approach to the de Sitter attractor as $A=B \sim
e^{\sqrt{\Lambda/3}\,t}$ and $P_B=2\,P_A \sim
e^{2\,\sqrt{\Lambda/3}\,t}$. One of the questions to be examined in this
paper is the characterization of sets of initial conditions for which
this asymptotic de Sitter attractor is attained.

Another important feature of the dynamical system (5) is to admit an
invariant manifold ${\cal{M}}$ defined by

\begin{equation}
A = B,\,\, P_A = P_B/2.
\end{equation}

On ${\cal{M}}$ the dynamics is governed by the two-dimensional system

\begin{equation}
\left\{
\begin{array}{ll}
\dot{B} = \frac{P_B}{8\,B} \\
\dot{P_B} = \frac{P_B^2}{16\,B^2} - 1 + 4\,\Lambda\,B^2\
\end{array}
\right. \
\end{equation}

\noindent The system (9) is integrable. Its phase portrait is depicted
Fig.  1, where the integral curves represent homogeneous and isotropic
universes with Bianchi IX compact spatial sections, with dust and a
cosmological constant.

We remark that the critical point $E$ belongs to ${\cal{M}}$. Contrary
to the models examined in Refs. \cite{calza} and \cite{cornish}, the
separatrices $S$ are present as solutions of the $full$ dynamical
equations.  As we will discuss in Sec. 3 and 4, the behavior of the
separatrices found in the previous studies will be played by the
cylinders emanating from unstable periodic orbits associated to the
saddle-center of $E$.

\section{Cylindrical Topology near the Critical Point $E$}

As a consequence of the saddle-center nature of the critical point $E$,
the dynamics of the models described by (5) exhibits completely new
features, which will be analyzed now. Part of this analysis is based on
Refs. \cite{osorio}\cite{vieira}\cite{wiggins}. The overall scenario
originates from both the anisotropy of the model and the cosmological
constant. To discuss the topology of the phase space in the neighborhood
of $E$, we make use of a theorem by Moser\cite{Moser} which states that
it is always possible to find a set of canonical variables such that, in
a small neighborhood of a saddle-center at the origin, the Hamiltonian
is expressed as

\begin{equation}
H(q_1,q_2,p_1,p_2)=\frac{\sqrt{\Lambda}}{2}\,(p_2^2-q_2^2) - \sqrt{2\,\Lambda}\,(p_1^2+q_1^2) + {\cal{O}}(3)-E_0 + E_{cr}=0.
\end{equation}

\noindent The critical point is located at the origin $q_1 = q_2 = p_1 =
p_1= 0$, with $E_0 = E_{cr}$. Here ${\cal{O}}(3)$ denotes non-quadratic
terms of the expansion and $\pm\,\sqrt{\Lambda}$,
$\pm\,2\,i\,\sqrt{2\,\Lambda}$ are the eigenvalues of the linearized
system about $E$.

Let us now restrict ourselves to a small neighborhood of $E$ such that
we can neglected ${\cal{O}}(3)$ and the quantity ${\cal{E}}=E_0-E_{cr}$
is small. The Hamiltonian reduces to

\begin{equation}
H(q_1,q_2,p_1,p_2) \sim \frac{\sqrt{\Lambda}}{2}\,(p_2^2-q_2^2) -\sqrt{2\,\Lambda}\,(p_1^2+q_1^2)
 - {\cal{E}} = 0.
\end{equation}

\noindent In this approximation, $H$ is separable with two approximate
constants of motion given by the partial energies

\begin{equation}
E_2=\frac{\sqrt{-\Lambda}}{2}\,(p_2^2-q_2^2),\,\, E_1=\sqrt{-2\,\Lambda}\,
(p_1^2+q_1^2)
\end{equation}

\noindent The energies $E_2$ and $E_1$ will be referred\cite{holmes} to
as the hyperbolic motion energy and the rotational motion energy of the
system about $E$, respectively. Note that $E_1$ is always positive. To
describe all possible motions, the following situations must be taken
into account.

If $E_2=0$ two possibilities arise. First, we have $p_2=q_2=0$ meaning
that the motions are unstable periodic orbits $\tau_{E_0}$ in the plane
$(p_1,q_1)$. Such orbits depend continuously on the parameter $E_1 \sim
-{\cal{E}}$ (cf. Fig. 2(a)). The second possibility will be
$p_2=\pm\,q_2$, which defines the linear stable $V_s$ and unstable $V_u$
$1$-dimensional manifolds of Fig. 2(b). These manifolds are tangent at
$E$ to the separatrices $S$ of the invariant manifold (described as
$q_1=0$, $p_1=0$ in the new variables) of Fig. 1. The separatrices are
actually the non-linear extension of $V_u$ and $V_s$. The direct product
of the periodic orbit $\tau_{E_0}$ with $V_s$ and $V_u$ generates, in
the linear neighborhood of $E$, the structure of stable ($\tau_{E_0}
\times V_s$) and unstable cylinders ($\tau_{E_0} \times V_u$), which
coalesce into the orbit $\tau_{E_0}$ for times going to $+\infty$ or
$-\infty$, respectively (cf. Figs. 2(c), 2(d)). The energy of any orbit
on these cylinders is the same of the periodic orbit $\tau_{E_0}$. It
can be showed that, in the non-linear regime (when non-quadratic terms
of the Hamiltonian must be taken into account), the plane-$(q_1,p_1)$ of
the rotational motion in the linear regime extends to a 2-dimensional
manifold - the {\it center manifold}\cite{simo} - of unstable periodic
orbits of the system. The intersection of the center manifold with the
energy surface

\begin{equation}
H({\cal{E}})=0
\end{equation}

\noindent (cf Eq. (11)) is a periodic orbit parametrized with $E_0$,
from which a pair of cylinders emanate. We can see easily from Eq. (11)
that the intersection of the central manifold with the energy surface
${\cal{E}}=0$ is just the point $E$. For ${\cal{E}}>0$ the energy
surface does not intersect the central manifold, the occurrence of the
structure of cylinders being therefore restricted to the energy surfaces
in which ${\cal{E}}<0$.

For the case $E_2 \neq 0$ and ${\cal{E}}<0$, the motion is restricted on
infinite cylinders resulting from the direct product of the hyperbolae
lying in the regions $I$ and $II$ of Fig. 2(b), with periodic orbits of
the central manifold in a small neighborhood of $E$. A general orbit
which visits the neighborhood of $E$ belongs to the general case $E_1
\neq E_2 \neq 0$. In this region the orbit has an oscillatory approach
to the linear cylinders (cf. Fig.  3), the closer as $E_2 \rightarrow
0$. For instance, the outcome of this oscillatory regime will be
collapse if $E_2<0$ or escape to de Sitter attractor if $E_2>0$ for
initial conditions taken in the quadrant $q_2<0$, $p_2>0$. This is for
the linear regime. In general, for orbits which visit a neighbohood of
$E$, the non-integrability of the Hamiltonian system (5) induces that
the partition of the energy ${\cal{E}}$ into the rotational mode energy
$E_1$ and the hyperbolic mode energy $E_2$ is chaotic. In another words,
given a general initial condition of energy ${\cal{E}}$, we are no
longer able to foretell which amount of ${\cal{E}}$ goes to each mode,
namely, in what of the regions $I$ or $II$ (Fig. 2(b)) the orbit will
land when approaches of $E$. This manifestation of the non-integrability
will be physically meaningful to characterizing a chaotic exit to
inflation, as we will show in the next section.

We must finally discuss the extension of the structure of cylinders
outside the neighborhood of $E$. Let us consider a linearized cylinder
associated to a periodic orbit with energy $E_0$ such that
${{\cal{E}}}=E_0-E_{cr}$ is very small. One possible way to examine the
extension of this cylinder is to consider the linearization of the
dynamical system (5) about the separatrix $S$. The equations for the
separatrix are denoted by

\begin{equation}
(a_S(t),p_S(t)),
\end{equation}

\noindent and consider the expansion about $S$

\begin{equation}
\left\{
\begin{array}{ll}
A = a_S+X,  \\ P_A = p_S + Z \\
B = a_S + Y, \\ P_B = 2\,p_S + W.
\end{array}
\right. \
\end{equation}

\noindent To the first order in $(X, Y, Z, W)$, the system (5) results

\begin{equation}
\left(
\begin{array}{c}
\dot{X}  \\ \dot{Y} \\ \dot{Z}  \\ \dot{W}
\end{array}
\right) \ = {\cal{A}}(t)\,\left(
\begin{array}{c}
X  \\ Y \\ Z  \\ W
\end{array}
\right)
\end{equation}

\noindent where ${\cal{A}}(t)$ is a $4 \times 4$ time-dependent matrix,
the entries of which are simple functions of $a_S$ and $p_S$. For the
sake of completeness we give its expression in the Appendix, but for our
purposes here it will be enough to remark that ${\cal{A}}(t)$ is
bounded, except for $a_S \rightarrow 0$. A matrix is defined to be
bounded is all its entries as well as its determinant are
bounded\footnote{A similar analysis can be made of the motion around
other integral curves of the invariant manifold depicted in Fig. 1. In
Eqs. (14) and (15) it is sufficient to substitute $a_S(t)$ and $p_S(t)$
by the solution of the integral curve $(a_I(t), p_I(t))$ associated to a
given energy $E_I$ in the invariant manifold.}.

Clearly, for a small neighborhood of $E$, the separatrix is approximated
by $V_s$ and $V_u$, and the linear cylinder based on $\tau_{E_0}$ will
be a solution of (16) by construction\cite{coddington}. Furthermore, it
follows from the system (16) that its extension will be contained in a
4-dimensional small tube about the separatrix, as far as $a_S(t)$ does
not go to zero. When $a_S(t) \rightarrow 0$ the above linear
approximation is no longer valid. Higher order terms become important
for the dynamics, and the non-integrability of the system results in the
distortion and twisting of the cylinders. The stable cylinder and the
unstable one will cross each other, producing chaotic
sets\cite{osorio}\cite{wiggins} and consequent "destroyed" regions of
the Poincar\'{e} maps of the system, as showed in Section 5. This also
occurs for cylinders emanating from periodic orbits of the center
manifold which are not in a small neighborhood of $E$ (periodic orbits
in the non-linear regime).

Finally, with view to the next section, let us select a 4-dimensional
small sphere of initial conditions about one point $S_0$ of the
separatrix with radius $R(S_0)$ of the order of the linear perturbation
in (15), namely, $R(S_0) \sim (X^2+Y^2+Z^2+W^2)^{1/2}$ (cf. Fig. 4). The
energy surfaces which intersects the sphere are those with $E_0$ in the
domain $|E_0-E_{cr}| \leq R(S_0)$. The cylinders (associated to the
periodic orbits $\tau_{E_0}$) in these surfaces will obviously intersect
the sphere. As we discussed before, if we evolve the sphere back $a_S
\rightarrow 0$ the small tube spread and twist, as well as the cylinder
associated to $\tau_{E_0}$. This stable cylinder and the unstable one
will cross each other and return eventually several times to the sphere.
This geometry will be the basis of the numerical experiments of the next
section. By evolving dynamically the sphere towards the neighborhood of
$E$, we show that for a band of energy, the sphere contains a chaotic
set which induces a chaotic exit to inflation.

\section{Chaotic Exit to Inflation}

In the numerical experiments performed here, we use the variables the
variables $(A, B, P_A, P_B)$. All calculations were made using the
package {\it Poincar\'{e}}\cite{ed} in a IBM compatible PC Pentium 133
with 64 MB of RAM memory, and a Fortran program to construct
Poincar\'{e} surface of sections, where we enforce that the error of the
Hamiltonian never exceed a given threshold of $10^{-10}$. Henceforth, we
assume $\Lambda=0.25$, so that the critical point $E$ is characterized
by $A=B=1.0, P_A=P_B=0$ and $E_{cr}=1$.

he phase space under consideration is not compact, and we will actually
identify a chaotic behavior associated to the possible asymptotic
outcomes of the orbits in this phase space, namely, escape to de Sitter
state attractor at infinity (inflationary regime) or collapse after a
burst of initial expansion. To begin with our numerical experiments, we
consider the invariant manifold ${\cal{M}}$ ($A=B$, $P_A=P_B/2$). The
basic characteristic of the curves in ${\cal{M}}$ have been discussed
already (cf. Fig. 1). We remark, however, that the separatrices define
the regions of collapse and expansion in ${\cal{M}}$, but not in the
$full$ 4-dimensional phase space. Following the theoretical backgrounds
presented in Sec. 3, we investigate numerically the behavior of orbits
in a domain near the separatrices. In essence, the sets of initial
conditions are constructed in the following way. Let $S_0$ be a point
belonging to the separatrix ($E_0=1.0$), with coordinates $A=B=0.4$,
$P_B=2\,P_A=1.357645019878171$. We are interested, therefore, in those
orbits representing expanding models after the initial singularity.
Around this point, we construct a 4-dimensional sphere in the phase
space with arbitrary small radius $R$, for instance $R=10^{-2}, 10^{-3},
10^{-4}$, etc (cf. Fig. 4). The values of $A$, $B$, $P_A$ and $P_B$ are
taken in energy surfaces which have a non-empty intersection with this
sphere as evaluated from the Hamiltonian constraint. As we have seen,
such energy surfaces are those for which the range of energy,
$\Delta\,E_0$, about $E_0=E_{cr}=1.0$ is of the order of, or smaller
than the radius $R$. On physical grounds, we are considering the
evolution of cosmological models with small anisotropic perturbations.

After several numerical experiments with the above sets of initial
conditions, we note that, as expected, two possible outcomes arise:
collapse or expansion into de Sitter configuration. The energy $E_0$
that varies from $1.0-\Delta\,E_0$ to $1.0+\Delta\,E_0$, with
$\Delta\,E_0 \sim R$, determines the long-time behavior of the orbits.
In Figs. (5) and (6), collapse and escape of 50 orbits are displayed for
$R=10^{-4}$. The orbits oscillate around the separatrix and approach the
critical point $E$ inside a sphere about this point with radius of the
same order of $R$. In this region, the linear approximation is valid and
the local separation of the dynamics into rotational motion and
hyperbolic motion (cf. Eq. 10) can be used to understand the results.
The final state of the orbits depends crucially on the partition of the
total energy ${\cal{E}}=E_0-E_{cr}$ into the rotational motion mode and
the hyperbolic motion mode. Hence, if $E_2>0$ the orbits escape, whereas
collapse is characterized by $E_2<0$. The rotational motion (energy
$E_1$) describes the oscillatory character of the orbits around the
critical point, indicating that some orbits can spend more time around
$E$ than others.

The main result of this section is to show that, for a determined
interval of energy $\delta\,E^*$ contained in the domain
${\cal{D}}=[1.0-\Delta\,E_0$, $1.0+\Delta\,E_0]$, this partition is
chaotic. Indeed, let us try to determine in ${\cal{D}}$ the values of
energy $E_0=E_{min}$ for which all orbits escape and $E_0=E_{max}$ for
which all orbits collapse. According to our numerical work, we find out
that for each sphere of initial conditions, there always exists a
non-null interval $\delta\,E^*=|E_{max}-E_{min}|$, where part of the
orbits escapes and another part collapses, resulting in an {\it
indeterminate} outcome. In Fig. (7), this behavior is showed for spheres
of initial conditions of radius $R=10^{-3}, 10^{-4}$ and $10^{-5}$. An
empirical relation between the gap $\delta\,E^*$ and the radius $R$ is
obtained and showed in Table 1. For spheres with $R \leq 10^{-2}$, we
have $\delta\,E^* \propto R^2$. Indeterminate outcome due to
$\delta\,E^* \neq 0$ occurs only for ${\cal{E}}=E_0-E_{cr}<0$, as
expected (indeed, if ${\cal{E}} \geq 0$ all orbits escape since, from
Eq. (11), $E_2>0$). The orbits collapse or escape, depending on the
partition of the energy ${\cal{E}}$ into the modes $E_1$ and $E_2$ (cf.
Eqs. 11 and 12), such that the later assumes negatives or posititves
values, respectively. The way in which this partition works, once a set
of orbits approach $E$, is completely unknown for energies in within the
gap $\delta\,E^*$. As a consequence, any infinitesimal fluctuation from
a given initial condition inside the sphere can lead to an indeterminate
outcome, that is, collapse or escape. This is the evidence of chaos in a
physically relevant context. In other words, we may state that the
boundaries of initial conditions for collapse and inflation are mixed as
a consequence of the crossing of cylinders.

The presence of chaos in the system is a consequence of the crossing of
stable and unstable cylinders, emanating from the unstable periodic
orbits of the center manifold, as dicussed before. This topological
structure is actually an invariant characterization of chaos. Finally,
we remark that the above behavior is not restricted to initial
conditions taken in small neighborhoods of points of the separatrix. Any
sets of initial conditions taken in an arbitrary neighborhood of the
invariant manifold ${\cal{M}}$ which result in orbits that visit a small
neighborhood of $E$, display the above chaotic behavior.

In a less simple model, radiation could be taken into account as a more
plausible matter field emerging after the Planck era\cite{hawking}
instead of dust. The effect of radiation is considered if we set the
equation of state as $p=\frac{1}{3}\,\rho$. The saddle-center critical
point and the invariant manifold are also present, and the same type of
behavior occurs if we construct sets of initial conditions as before.
There is also a similar relation between the gap $\delta\,E^*$ and the
radius $R$ showed in the Table 2. A more complete and detailed analysis
will be subject of a forthcoming paper.

\section{Further Numerical Results and Wald's Phenomenon}

We consider here the evolution of completely anisotropic models taking
initial conditions far from the invariant manifold ${\cal{M}}$. It is a
remarkable fact that the Hamiltonian (4) is regular at the plane $A=0$
where the curvature is singular. Therefore, from the point of view of
the Hamiltonian dynamics, orbits can be analitycally extended beyond the
singularity to the domain $A<0$. Although these orbits are not
physically meaningful in this domain, they are nevertheless essential in
the description of the underlying geometrical structure of the full
Hamiltonian system and its non-integrability. In this context we will
give a geometrical picture for Wald's inequality\cite{wald}.

In Fig. 8(a), we exhibit the Poincar\'{e} map of the system for
$(E_0=0.92504831163435, \Lambda=0.25)$ with surface of section $(P_B=0,
\dot{P_B}>0)$. This map makes explicit the coexistence of KAM tori and
destroyed regions in the phase space, as a consequence of the
non-integrability of the system. We were not able to find a torus
totally contained in the domain $A>0$. Thus, orbits on, or inside the
tori (with initial conditions taken on $A>0$) necessarily collapse. This
can be seen from double Poincar\'{e} map with surface of section $P_B=0$
as showed in Fig. 8(b), where $(E_0=0.15, \Lambda=2.0$). On the
contrary, orbits in the destroyed region are free to escape. Orbits on a
torus with analytical continuation to a domain of negative $A$ are said
to perform cosmic cycles (according with Ref. \cite{calza}).

Our next step is to examine the effect of varying the value of the
cosmological constant $\Lambda$ on the tori structure in the same region
of phase space. As can be verified numerically, the effective result of
increasing $\Lambda$ is the destruction of the tori. We showed this in
Fig. 9, where orbits are drawn in the same region of initial conditions
but with distinct values of $\Lambda$. If we change $\Lambda$ from
$0.25$ to $0.30$ (the energy changes from $E_0=0.8$, taken initially, to
$E_0=0.8000050001250895$), orbits initially in a torus escape to the de
Sitter attractor after some cycles; if we further increase $\Lambda$, no
cosmic cycles takes place before escape. The phenomenon of destruction
of tori, by increasing the value of the cosmological constant, is a
geometrical picture for Wald's inequality. We nevertheless remark that,
for any value of the cosmological constant (however large, except
infinity), there will always be collapsing orbits.

\section{Final Remarks and Conclusions}

In this paper we have discussed the dynamics of Bianchi-IX models, which
may provide a description of pre-inflationary stages of the universe.
The main ingredients of the models are a cosmological constant, arising
as the vacuum of the inflaton field, and anisotropy. The degrees of
freedom are taken in the gravitational sector, and for simplicity we
restrict ourselves to two distinct scale factors only. The presence of
anisotropy, even in the form of small perturbations, together with the
cosmological constant produces in the phase space of the system a
critical point identified as a saddle-center. Associated to the
saddle-center we have a 2-dimensional manifold of unstable periodic
orbits, and the structure of infinite unstable cylinders emanating from
them. The stable and unstable cylinders coalesce into the corresponding
periodic orbit for time going to $\infty$ and $-\infty$, respectively.
The non-integrability of the system results in the distortion and
twisting of the cylinders, and the eventual intersection of one (stable)
with the other (unstable) with a consequent chaotic behavior  of the
dynamics of the phase space. This behavior of the cylinders is analogous
to the one played by the separatrices (heteroclinic curves connecting
saddle points) of the gravitational sector in the models discussed in
\cite{calza} and \cite{cornish}. In these models the gravitational
variable dynamics couples with the scalar fields giving rise in the
heteroclinic breaking and tangle which is the origin of chaos in these
models.

The results of Section 4, where we describe chaotic exit to inflation,
extends the result of Cornish and Levin\cite{cornish} for the case of
two gravitational degrees of freedom. The breaking of the boundary
between initial conditions domains of collapse and escape to inflation,
showed in \cite{cornish}, is indeed meaningful for the case of one
gravitational degree of freedom only (the unperturbed separatrix is in
fact a sharp boundary between the domains of initial conditions). In our
case, the separatrices are present in the full dynamics and define
sharply regions of collapse and inflation in the invariant manifold, but
not in the full 4-dimensional phase space  of the gravitational
dynamics. However, for each small sphere of initial conditions taken
about one point of the separatrix, it is always possible to find a small
domain (or gap) of energy such that the intersection of the sphere with
an energy surface associated to the above domain is a chaotic set in the
sense of discussed in Section 4. Namely, a small perturbation in initial
conditions taken in this set would change  an orbit from collapse to
escape to the de Sitter phase. Furthermore, we conjecture that it is
feasible to select initial conditions in this set such that the scale
factors oscillate an arbitrary fixed time $T$ about $E$ ($T=\infty$
included) before collapsing or escaping to a de Sitter phase. Putting in
another words, chaos is established by the incertainty in the  partition
of the energy $E_0$ into the rotational motion energy and hyperbolic
motion energy about $E$.

The oscillations of the scale factors about $E$ may have an important
physical effect concerning inhomogeneous perturbations. Let them be
scalar field perturbations and/or matter density perturbations in this
gravitational background. In fact, the time-dependent amplitude of each
Fourier component of the perturbation will satisfy a linear differential
equation, the coefficients of which will be oscillatory functions about
$E$. As we have seen, the latter may be approximated by a periodic
function describing $\tau_{E_0}$. Therefore, by a resonance mechanism,
there will occur amplification of the particular Fourier components
having period approximately equal to the period $\tau_{E_0}$ (we
obviously assume here that no dissipation effects are included in the
equations of motion of the perturbation in this stage). Even if the
universe inflates afterwards, the relative ratio of amplitudes produced
after this mechanism of amplification will be nevertheless mantained.

Finally, as discussed in Section 4, we have given a geometric
interpretation for  Wald's result in terms of invariant tori and their
destruction by increasing of the cosmological constant. For a given
value of the cosmological constant, the phase space presents the
structure of KAM tori, and orbits on, or inside these tori collapse
necessarilly. However, if we increase the value of the cosmological
constant, the tori are destroyed and orbits previously on, or inside
them may eventulally escape to de Sitter configuration.

\section{Acknowledgements}

The authors are grateful to CNPq for financial support.

\section{Appendix}

The matrix ${\cal{A}}(t)$ is given by

$${\cal{A}} = \left(\begin{array}{cccc}
		-\frac{p_S}{4\,a_S^2} & 0 & -\frac{1}{4\,a_S} & \frac{1}{4\,a_S}  \\
		0  & -\frac{p_S}{4\,a_S^2} & \frac{1}{4\,a_S} & 0  \\
		\frac{5}{a_S}  &  -(\frac{3}{a_S}+\frac{p_S^2}{4\,a_S^3}+4\,\Lambda\,a_S) & 
		\frac{p_S}{4\,a_S^2} & 0   \\
		-(\frac{3}{a_S}+\frac{p_S^2}{4\,a_S^3}+4\,\Lambda\,a_S)  & 
(\frac{3}{a_S}+\frac{p_S^2}{4\,a_S^3}-4\,\Lambda\,a_S) & 0 & 
\frac{p_S}{4\,a_S^2}
	    \end{array} \right)$$

\newpage

\section*{Figure Captions}

Fig. 1 Integral curves on the invariant manifold $A=B$, $P_A=P_B/2$. The
orbits on regions $(I)$ have $ E_0 < \frac{1}{\sqrt{4\,\Lambda}}$,
whereas for those orbits on $(II)$, $E_0 > \frac{1}{\sqrt{4\,\Lambda}}$.
The separatrices are characterized by the energy $E_0=E_{cr}
=\frac{1}{\sqrt{-4\,\Lambda}}$.

Fig. 2 (a) Periodic orbit of the Hamiltonian system (5) in the linear
approximation and projected onto plane-$(q_1, p_1)$ of the normal
variables. (b) The linear unstable $V_u$ and stable $V_s$ 1-dimensional
manifolds. The hyperbolae are the linearized solutions in the plane $(B,
P_B)$ of the saddle for $E_2<0$ (region $I$) and $E_2>0$ (region $II$).
(c) Stable and unstable cylinders manifolds emanating from the periodic
orbit $\tau$. They are the non-linear extension of the linearized
cylinders $\tau \times V_u$ and $\tau \times V_s$ in the neighborhood of
$E$. (d) Numerical ilustration of the linear stable and unstable
cylinders for $\Lambda=0.25$ and $E_0=0.9999999$ in the neighborhood of
$E$.

Fig. 3 General orbits with oscillatory approach to the cylinders on the
neighborhood of $E$ corresponding to $E_0=0.99999999$. 
on the plane $(q_2, p_2)$ of the above orbits. Note the 
asymptotes $V_u$ and $V_s$.

Fig. 4 Projection of the 4-dimensional sphere of initial conditions
about the point $S_0$ of the separatrix $S$.

Fig. 5 Collapse of 50 orbits initially inside an sphere of initial
conditions of radius $R=10^{-4}$ for $E_0=0.9999999937$ and
$\Lambda=0.25$. (a) View of orbits projected onto the plane-$(A, P_A)$.
(b) Zoom of the region near the critical point. Note the oscillations
around the separatrix as well as the critical point $E$.

Fig. 6 Escape of 50 orbits to the de Sitter configuration. We consider
an sphere of initial conditions of radius $R=10^{-4}$ within an energy
surface $E_0=0.99999999999$.  (a) View of orbits projected onto the
plane-$(A, P_A)$. (b) Zoom of the tridimensional region near the
critical point.

Fig. 7 Chaotic exit to inflation: (a) outcome  of 50 orbits showed in
the plane-$(A, P_B)$ for $E_0=0.99999999771$ and the sphere of initial
conditions with radius $R=10^{-4}$. (b) Tridimensionl view of the region
close to the critical point $E$ for $R=10^{-3}$. (c) View of the
plane-$(A, P_B)$ for the case $R=10^{-5}$ and energy
$E_0=0.999999999939$. Note that the orbits remains in the neigbhorhood
of $E$ of same order of the initial sphere.

Fig. 8 (a) Poincar\'{e} surface of section $P_B=0, \dot{P_B}>0$ for
$\Lambda=0.25$ and $E_0=0.92504831163435$. (b) Double Poincar\'{e}
surface of section with the condition $\dot{P_B}>0$ relaxed with
$E_0=0.15$ and $\Lambda=2.0$. It is clear the presence of destroyed
regions of the phase space together with KAM tori structure. The tori
cross the plane $A=0$ performing the so called cosmic cycles.

Fig. 9 (a) Two collapsing orbits inside a torus with $\Lambda=0.25$ and
$E_0$. (b) The result of increasing the cosmological constant to
$\Lambda=0.30$, keeping the same region of the phase space and assuming
$E_0=0.8000050001250895$, is the destruction of the torus: after some
cycles, both orbits escape to de Sitter configuration.

Table 1

Relation between the radius $R$ and the gap of energy $\delta\,E_0^*$
for the case $\gamma=0$ (dust).

Table 2

Relation between the radius $R$ and the gap of energy $\delta\,E_0^*$
for the case $\gamma=1/3$ (radiation).

\begin{tabular}{|c|c|}
\hline
$R$ & $\delta E_0^*$ \\[2mm]
\hline
\hline
$10^{-2}$ & $0.928 \times 10^{-4}$ \\[2mm]
\hline
$10^{-3}$ & $0.618 \times 10^{-6}$ \\[2mm]
\hline
$10^{-4}$ & $0.617 \times 10^{-8}$ \\[2mm]
\hline
$10^{-5}$ & $0.619 \times 10^{-10}$ \\[2mm]
\hline
$10^{-6}$ & $0.630 \times 10^{-12}$ \\[2mm]
\hline
\hline
\end{tabular}

\vspace{0.5 cm}

Table 1

\vspace{1.0 cm}

\begin{tabular}{|c|c|}
\hline
$R$ & $\delta E_0^*$ \\[2mm]
\hline
\hline
$10^{-2}$ & $2.874 \times 10^{-4}$ \\[2mm]
\hline
$10^{-3}$ & $2.890 \times 10^{-6}$ \\[2mm]
\hline
$10^{-4}$ & $2.855 \times 10^{-8}$ \\[2mm]
\hline
$10^{-5}$ & $2.880 \times 10^{-10}$ \\[2mm]
\hline
$10^{-6}$ & $2.830 \times 10^{-12}$ \\[2mm]
\hline
\hline
\end{tabular}

\vspace{0.5 cm}

Table 2

\end{document}